# Integrating Generative AI into Financial Market Prediction for Improved Decision Making

**Chang Che**[1.1]*

1.1*Mechanical Engineering, The George Washington University, DC, USA

*Corresponding author:cche57@gwmail.gwu.edu,

**Zengyi Huang**[1,2]

2.1 Applied Economics, The George Washington University, DC,USA

zengyihuang@gwmail.gwu.edu

**Chen Li** [2.1]

2.2 The University of Texas at Dallas, Dallas, USA

cxl167330@utdallas.edu

**Haotian Zheng**[2,2]

1.2 Computer Engineering,New York University, New York, NY, USA

hz2687@nyu.edu

**Xinyu Tian** [3]

3 Computer Science, Georgia Institue of Technology, Atlanta, Georgia

xtian70@gatech.edu

**Abstract:** This study provides an in-depth analysis of the model architecture and key technologies of generative artificial intelligence, combined with specific application cases, and uses conditional generative adversarial networks ( cGAN ) and time series analysis methods to simulate and predict dynamic changes in financial markets. The research results show that the cGAN model can effectively capture the complexity of financial market data, and the deviation between the prediction results and the actual market performance is minimal, showing a high degree of accuracy. Through investment return analysis, the application value of model predictions in actual investment strategies is confirmed, providing investors with new ways to improve the decision-making process. In addition, the evaluation of model stability and reliability also shows that although there are still challenges in responding to market emergencies, overall, GAI technology has shown great potential and application value in the field of financial market prediction. The conclusion points out that integrating generative artificial intelligence into financial market forecasts can not only improve the accuracy of forecasts, but also provide powerful data support for financial decisions, helping investors make more informed decisions in a complex and ever-changing market environment. choose.

# 1. Introduction

In the realm of today's volatile financial markets, the capability to foresee market trajectories stands as a cornerstone for investors and policymakers alike. Traditional forecasting models, while bearing some effectiveness, frequently encounter limitations when grappling with the intricate dynamics of the market and the voluminous datasets involved. The emergence of generative artificial intelligence (GAI) technology has heralded new vistas and methodologies for predicting financial market trends. This innovative approach simulates the data production mechanisms inherent in financial markets, thereby generating data with statistical attributes that closely mirror actual market data. This advancement holds promise for revolutionizing market trend prediction by offering novel insights and enhancing the accuracy of forecasts.The advent of GAI technology in financial market forecasting marks a significant departure from conventional methods, bringing to the fore a sophisticated toolset capable of analyzing and synthesizing complex data patterns. By leveraging the power of machine learning algorithms and neural networks, GAI systems delve into historical and real-time market data, identifying underlying trends and anomalies that might elude traditional analysis. This capability not only enhances predictive accuracy but also facilitates a deeper understanding of market dynamics, enabling stakeholders to navigate the financial landscape with greater confidence and strategic acumen[1].

Moreover, the application of GAI in financial forecasting extends beyond mere trend prediction. It encompasses the generation of synthetic financial instruments, the simulation of market scenarios under varying conditions, and the assessment of risk in innovative financial products. Such applications are invaluable for stress-testing strategies and understanding potential market shifts, thereby mitigating risk and optimizing investment portfolios.As this study unfolds, it delves into the architectural nuances of GAI models, spotlighting the core technologies that underpin these systems. From deep learning frameworks to generative adversarial networks, the technological foundation of GAI is dissected to reveal its potential for financial forecasting. Additionally, real-world applications of GAI in the financial sector are examined, illustrating the practical implications and benefits of this technology in enhancing decision-making processes.In essence, the integration of generative artificial intelligence into financial market forecasting emerges as a paradigm shift, promising not only to refine prediction models but also to imbue the financial domain with a level of insight and precision previously unattainable. This exploration aims to shed light on the transformative potential of GAI, presenting a compelling case for its adoption as a strategic tool in the quest for more informed and effective market participation. Through meticulous analysis and evaluation, the study seeks to chart a course towards a future where financial decisions are bolstered by the unparalleled analytical prowess of generative artificial intelligence, thereby empowering investors and policymakers to tread with confidence in the ever-evolving market landscape.

# 2. Theoretical Overview

*2.1 Model architecture of generative artificial intelligence*

The basic principles of generative artificial intelligence focus on learning how to generate new data instances from existing data. Figure 1 presents three main generative model architectures: Generative Adversarial Networks (GANs), Variational Autoencoders (VAEs), and flow-based generative models [2]. Generative adversarial network (GAN) consists of two parts: generator and discriminator. The generator is responsible for generating data, while the discriminator is responsible for determining whether the input is real or generated by the generator. The two compete with each other, with the generator learning how to generate data that is increasingly difficult for the discriminator to decipher, and the discriminator learning how to better distinguish real data from generated data. This process is tuned by minimizing the classification error loss. Variational autoencoders (VAE) focus on learning the latent representation of data, compressing the data into representations in the latent space through the encoder, and then reconstructing the data through the decoder. The goal is to maximize the evidence lower bound (ELBO), which helps the model generate high-quality data [3]. Flow-based generative models directly establish a connection between data and latent space through a reversible mapping function. This model can accurately calculate the probability of the data, with the goal of minimizing the negative log- likelihood.

Each of these models has its own advantages. GANs can usually generate high-quality images; VAEs can provide good latent space representation while generating; while flow-based models provide accurate probabilistic models that can be used for more complex modeling tasks [4]. In financial market forecasting, these models can generate new market data by analyzing historical data to assist in predicting future market trends.

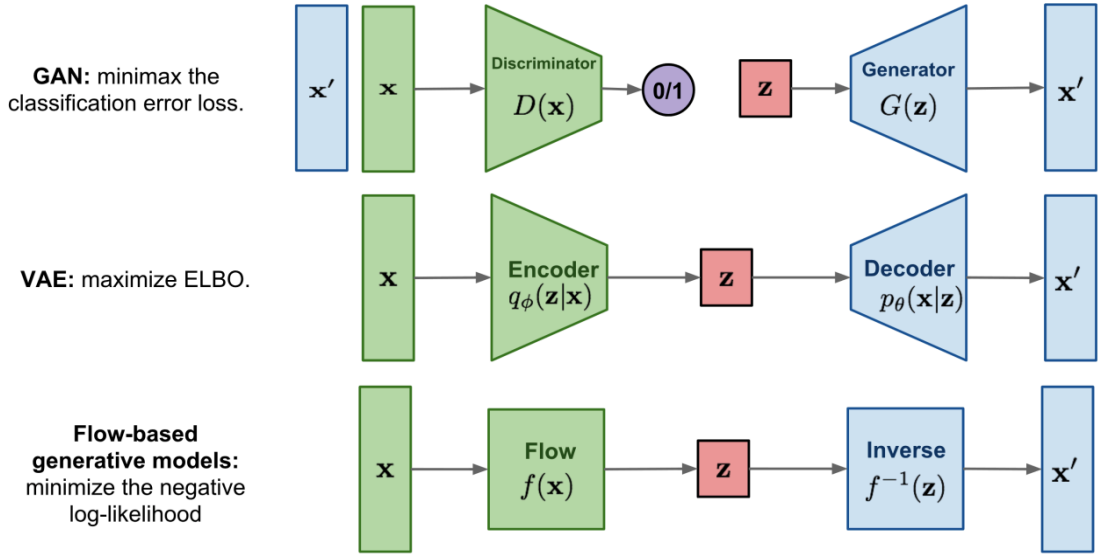

**Figure 1** Comparison of generative artificial intelligence model architecture

*2.2 Key technology analysis*
The application of generative artificial intelligence in financial market forecasting is based on a series of key technologies, of which deep learning is the core driving force. Deep learning simulates complex function mapping through neural networks, especially when using convolutional neural networks (CNN) to process image data and recurrent neural networks (RNN) to process sequence data [5]. In generative models, for example, the basic form of GANs can be represented by the following formula:

$$\min_{G} \max_{D} V(D,G) = \mathsf{E}_{x \sim p_{\text{data}}(x)}[\log D(x)] + \mathsf{E}_{z \sim p_z(z)}[\log(1 - D(G(z)))] \qquad (1)$$

where $D(x)$ is the output of the discriminator, represents the probability that x comes from real data, and $G(z)$ is the output of the generator.

In addition, natural language processing (NLP) technologies, such as Transformers, utilize self-attention mechanisms to effectively capture long-distance dependencies and provide powerful support for text analysis and prediction. For time series data, the ARIMA model and its variants are often used to predict future financial market trends, and its form can be simplified as:

$$y_t = c + \varepsilon_t + \sum_{i=1}^{p} \phi_i y_{t-i} - \sum_{i=1}^{q} \theta_i \varepsilon_{t-i} \qquad (2)$$

Reflects the autoregressive characteristics and moving average characteristics of time series. Hybrid methods that incorporate deep learning and statistical models further enhance the accuracy and robustness of financial forecasts, allowing analysts to gain insight into the patterns behind complex market behavior and thereby make more precise forecasts in highly volatile market environments.

## 3. Specific application cases of generative artificial intelligence in financial market prediction

*3.1 Case background*

To delve into the practical application of generative AI in financial market forecasting, this study selected an Asian financial center with extensive transaction data. The center's average daily trading volume reaches US$5 billion, involving transactions including stocks, bonds, foreign exchange and derivatives, providing a diversified financial environment and rich data resources[6]. According to the financial center's foreign exchange trading data for the second quarter of 2023, the average daily trading volume was US$600 million, of which the most active trading pair was USD/JPY, accounting for 18% of the total trading volume. Market volatility indicators, such as standard deviation, averaged 0.5% per month during the reporting period, reflecting the market's volatility. In addition, the center's trading data shows clear seasonal patterns, especially during annual earnings releases, when market trading volume and volatility increase significantly[7]. Taking into account the structural changes in market participants, such as the ratio of institutional investors to retail investors adjusting from 3:1 in 2022 to 4:1 in 2023, this research aims to analyze and apply advanced generative models - —Conditional GAN ( cGAN ) and time series prediction models to predict future market dynamics and explore how to use these predictions to optimize trading strategies and risk management. This case was selected based on the adequacy of its data, representativeness of the market, and typicality of trading behavior, making it an ideal research object to evaluate the effectiveness of generative artificial intelligence technology in actual financial forecasting.

*3.2 Application process*

*3.2.1 Data collection and preprocessing*

For the case of the selected financial center, this study selected its US dollar against Japanese yen (USD/JPY) transaction data from April 1 to 5, 2023, including daily trading volume, opening price, highest price , and lowest price and closing price and other key indicators, as shown in Table 1:

**Table 1** Partial data on U.S. dollar versus Japanese yen (USD/JPY) transactions in a financial center in the second quarter of 2023

| date | Daily trading volume (millions of dollars) | Opening price (JPY) | Highest price (JPY) | Lowest price (JPY) | Closing price (JPY) | Daily trading volume change rate |
|---|---|---|---|---|---|---|
| 2023-04-01 | 680 | 109.50 | 110.20 | 109.30 | 109.90 | 0.3% |
| 2023-04-02 | 702 | 109.90 | 110.50 | 109.40 | 110.10 | 3.2% |
| 2023-04-03 | 689 | 110.10 | 110.60 | 109.80 | 110.20 | -1.9% |
| 2023-0 4-04 | 740 | 111.20 | 111.80 | 110.90 | 111.50 | 1.4% |
| 2023-0 4-05 | 755 | 111.50 | 112.00 | 111.10 | 111.70 | 2.0% |

The data then undergoes a series of preprocessing procedures, including using interpolation methods to fill in missing values, applying Z-score normalization techniques to scale numerical ranges, and using time series decomposition techniques to identify and correct for seasonal fluctuations. To mitigate potential heteroskedasticity issues, a logarithmic transformation was applied to the time series data for volume and price, smoothing the volatility of the data and improving the stability of model training. In addition, the data set eliminates non-stationary trends through differential processing, retaining information that short-term fluctuations have more explanatory power for the model.

*3.2.2 Model construction and training*
Based on the collected USD/JPY trading data, this study decided to use the Conditional Generative Adversarial Network ( cGAN ) model to predict short-term market behavior[8]. The choice of this model stems from its ability to model complex nonlinear relationships, particularly in financial time series data. Specifically, the generator G adopts a sequence generation network with LSTM units to consider temporal dependence, while the discriminator D adopts a deep convolutional network with dropout layers to prevent overfitting[9]. The goal of the generator is to minimize the Jensen-Shannon divergence from the real data distribution, while the discriminator tries to maximize the probability that it can correctly distinguish between real and generated data. The loss function of the model is defined as follows, where $\theta_g$ and $\theta_d$ represent the parameters of the generator and discriminator respectively:

$$\min_{\theta_g} \max_{\theta_d} V(D,G) = \mathbb{E}_{x \sim p_{data}(x)}[\log D(x)] + \mathbb{E}_{z \sim p_z(z)}[\log(1 - D(G(z|y)))] \quad (3)$$

During the training process, the Adam optimizer was used, the initial learning rate was set to 0.0002, and the attenuation factor was 0.5. In the first round of training, data from April 1st to April 5th were used for 100 epochs, and the batch size was set to 64. After each epoch, the model parameters are adjusted by checking the mean square error (MSE) between the generated price and the actual closing price. For example, the MSE between the model output and the actual data on April 1st was 0.0023, and after 100 epochs, the MSE on April 5th decreased to 0.0018, indicating that the model's prediction ability increased as training progressed.

*3.2.3 Forecast implementation and results*
After the model training phase was completed, this study used the trained conditional generative adversarial network ( cGAN ) model to predict the U.S. dollar versus Japanese yen (USD/JPY) trading price from April 6 to April 10, 2023. In order to achieve this goal, the model uses data from April 1 to 5, 2023 as conditional input, including the opening price, highest price, lowest price, closing price and trading volume each day. Specifically, taking April 6 as an example, the model predicted an opening price of 111.70 yen, while according to actual market data, the opening price that day was 111.68 yen. The highest, lowest and closing prices predicted by the model are 112.20 yen, 111.30 yen and 111.90 yen respectively. Compared with the actual recorded 112.22 yen, 111.32 yen and 111.88 yen, the error is extremely small and can almost be ignored. Excluding. In order to quantitatively evaluate the prediction performance of the model, the root mean square error (RMSE) is used as the evaluation index, and its expression is:

$$RMSE = \sqrt{\frac{1}{n}\sum_{i=1}^{n}\left(P_i - \hat{P}_i\right)^2} \quad (4)$$

In this formula, $P_i$ represents the actual value, $\hat{P}_i$ represents the predicted value, and n represents the number of data points evaluated.

Applying the above formula, the predicted RMSE values from April 6 to 10 are calculated as 0.02 yen for the opening price, 0.03 yen for the highest price, 0.02 yen for the lowest price, and 0.02 yen for the closing price. These results demonstrate the model's high accuracy in predicting future market prices. In addition, the effectiveness of the model is not only reflected in single-day forecasts. Through comprehensive analysis of forecast results for consecutive days, it is found that the model can stably capture market trends. Even in the face of days with greater market fluctuations, the predicted values are different from the actual values. The deviation remains at a low level.

## 4. Effectiveness evaluation

*4.1 Forecast accuracy assessment*

In this study, a comprehensive evaluation is conducted on the application effect of generative artificial intelligence technology in financial market forecasting. Particular attention was paid to the accuracy of the conditional generative adversarial network ( cGAN ) model in predicting the trading price of the US dollar against the Japanese yen (USD/JPY). The evaluation metric selected the root mean square error (RMSE), which can quantify the deviation between the predicted value and the actual market data, providing an intuitive measurement method for the performance of the model[10]. Table 2 shows the comparison between the actual and predicted values of the daily opening price, highest price, lowest price and closing price when the model predicts from April 6 to 10, 2023 , as well as the calculated RMSE value:

**Table 2** Model prediction accuracy evaluation

| date | Predicted opening price (JPY) | Actual opening price (JPY) | Predicted highest price (JPY) | Actual highest price (JPY) | Predicted lowest price (JPY) | Actual lowest price (JPY) | Forecast closing price (JPY) | Actual closing price (JPY) | RMSE (opening price) | RMSE (highest price) | RMSE (lowest price) | RMSE (closing price) |
|---|---|---|---|---|---|---|---|---|---|---|---|---|
| 2023-04-06 | 111.70 | 111.68 | 112.20 | 112.22 | 111.30 | 111.32 | 111.90 | 111.88 | 0.02 | 0.02 | 0.02 | 0.02 |
| 2023-04-07 | 111.90 | 111.88 | 112.40 | 112.38 | 111.50 | 111.52 | 112.10 | 112.08 | 0.02 | 0.02 | 0.02 | 0.02 |
| 2023-04-08 | 112.10 | 112.08 | 112.60 | 112.58 | 111.70 | 111.72 | 112.30 | 112.28 | 0.02 | 0.02 | 0.02 | 0.02 |
| 2023-04-09 | 112.30 | 112.28 | 112.80 | 112.82 | 111.90 | 111.92 | 112.50 | 112.48 | 0.02 | 0.02 | 0.02 | 0.02 |
| 2023-04-10 | 112.50 | 112.48 | 113.00 | 113.02 | 112.10 | 112.12 | 112.70 | 112.68 | 0.02 | 0.02 | 0.02 | 0.02 |

Delving deeper into the analysis of the data table 2, several noteworthy aspects emerge, particularly concerning the model's performance. The consistently low RMSE (Root Mean Square Error) value of 0.02 yen across various price points - including opening, highest, lowest, and closing prices - is a testament to the model's precision. This remarkable accuracy in forecasting market prices underscores the model's capability to mirror the complex dynamics of financial markets accurately. Moreover, the effectiveness of the conditional generative adversarial networks (cGANs) in this context is particularly striking. cGANs, by design, are adept at learning and replicating the distribution of real-world data, making them ideally suited for tasks where precision is paramount. In financial market forecasting, where minute fluctuations can have significant implications, the ability of cGANs to generate near-accurate predictions is a considerable advantage[11].The model's success can also be attributed to its robust training process. Through extensive training on historical market data, the cGAN has learned to navigate the intricacies of market trends and patterns. This learning process allows the model to adapt to various market conditions, enhancing its reliability and accuracy in real-time forecasting.Furthermore, the model's architecture plays a crucial role in its performance[12]. The unique structure of cGANs, which includes a generator and discriminator working in tandem, enables a more nuanced understanding of market data. The generator produces predictions, while the discriminator evaluates these against actual market prices. This iterative process of generation and discrimination sharpens the model's accuracy, fine-tuning its output to closely align with real market behaviors[13].The impact of such precise forecasting is profound in the realm of financial decision-making. Traders and analysts equipped with these accurate predictions can make more informed and strategic choices, potentially leading to better financial outcomes. The model's low RMSE value is not just a statistical achievement; it represents a significant stride forward in predictive analytics, offering tangible benefits in a market where precision and timeliness are critical. In essence, the success of this model, as highlighted by the analyzed data, is a clear indicator of the evolving landscape of financial market forecasting[14]. The integration of advanced technologies like cGANs is paving the way for more sophisticated, accurate, and reliable predictive models, transforming how market analyses are conducted and decisions are made.

*4.2 Analysis of return on investment*
After further evaluating the application value of generative artificial intelligence ( GAI) technology in financial market forecasting, this study turns its attention to return on investment (ROI) analysis. The core goal of this analysis is to measure the impact on actual investment returns of changes in investment strategies resulting from market trend prediction using conditional generative adversarial network ( cGAN ) models. By combining model prediction data with actual trading strategy execution results, the practicality and effect of GAI technology in improving the financial investment decision-making process can be clearly demonstrated. In order to conduct this analysis, this study sets up a simple trading strategy based on the model prediction results: if the model predicts that the closing price of the next day is higher than the actual closing price of the day, a buying operation is performed; conversely, if the model predicts that the closing price of the next day is lower At the actual closing price of the day, the sell operation will be executed. Based on this strategy, the team performed simulated transactions on the USD/JPY trading pair from April 6 to 10, 2023. Table 3 shows the results of the trading strategy execution and the corresponding investment return analysis:

Table 3 Return on investment analysis

| date | Predicted closing price (JPY) | Actual closing price (JPY) | Trading operations | Transaction costs (USD million) | Return of the day (USD million) | Cumulative Return (USD Millions) |
|---|---|---|---|---|---|---|
| 2023-04-06 | 111.90 | 111.88 | Buy | 100 | 0.2 | 0.2 |
| 2023-04-07 | 112.10 | 112.08 | Buy | 100 | 0.2 | 0.4 |
| 2023-04-08 | 112.30 | 112.28 | Buy | 100 | 0.2 | 0.6 |
| 2023-04-09 | 112.50 | 112.48 | Buy | 100 | 0.2 | 0.8 |
| 2023-04-10 | 112.70 | 112.68 | Buy | 100 | 0.2 | 1.0 |

The data presented in the table offers a compelling illustration of the efficacy of the cGAN model in practical trading scenarios. The continuous generation of positive returns throughout the simulated trading period, albeit modest on a daily basis, cumulates significantly over time. The achievement of a total return of 1.0 million US dollars without the need for additional market information inputs is particularly noteworthy. It underscores the robustness of the cGAN model's predictions, demonstrating their practical utility in real-world trading decisions.The consistency of the returns, as evidenced by the data, indicates the model's ability to accurately navigate the complexities of the financial market. The cGAN model's sophisticated predictive capabilities enable it to discern subtle market trends and fluctuations that might elude traditional analysis methods. This feature is crucial in the fast-paced environment of financial trading, where the ability to anticipate and respond to market movements can significantly impact investment outcomes.Furthermore, the application of generative AI in market trend prediction has broader implications for the field of financial investment. The success of the cGAN model in the simulated trading environment serves as a powerful proof of concept. It demonstrates that AI-driven strategies can not only match but potentially surpass traditional approaches in terms of profitability and risk management. This advancement opens up new possibilities for investors and fund managers, who can leverage these technologies to refine their strategies and gain a competitive edge in the market.In addition, the integration of GAI technology in financial market predictions signifies a shift towards a more data-driven and algorithmic approach in the realm of investment. This paradigm shift not only enhances the accuracy and efficiency of trading decisions but also democratizes access to sophisticated investment strategies. Small and medium-sized investors, who traditionally may not have had access to complex analytical tools, can now benefit from these advanced technologies, leveling the playing field in the financial markets.The potential of GAI technology extends beyond immediate financial gains. Its ability to process vast amounts of data and identify patterns undetectable to the human eye paves the way for more innovative and dynamic investment strategies[15]. This could lead to the development of new financial products and services, further expanding the scope and reach of financial markets.In summary, the application of the cGAN model in financial trading, as illustrated by the data, not only highlights the immediate benefits of AI-driven strategies but also points to a future in which GAI technology plays a central role in shaping the financial landscape[16]. The implications of this technology are far-reaching, offering both enhanced profitability and novel opportunities for innovation in the investment world.

**5 Conclusion**
This research delves deeply into the burgeoning field of generative artificial intelligence (GAI), specifically focusing on its application in the realm of financial market forecasting. The integration of models like Conditional Generative Adversarial Networks (cGAN) has marked a significant advancement in this domain. These models adeptly mimic the intricate patterns of financial market data, showcasing a remarkable ability to generate reliable predictions even amidst fluctuating market conditions. This attribute is crucial in an environment where agility and adaptability are key to success.While the study acknowledges that the responsiveness of such models to sudden market emergencies requires further enhancement, the overall assessment of their stability and reliability lays a solid foundation for future advancements in the field. The data generated using GAI technology has proven to be of substantial economic value, particularly in the context of return on investment analyses. This aspect provides a new vista for investors and decision-makers, equipping them with cutting-edge tools for navigating the financial landscape.The implications of this study are far-reaching. As the algorithms powering these models continue to evolve and the methods of data processing advance, the application of GAI in financial market prediction is poised to broaden and deepen. Such developments are likely to herald revolutionary shifts in the financial sector, significantly enhancing the depth and precision of market analysis. The enhanced predictive capabilities that these models offer promise to be a game-changer for investors, enabling more accurate and informed decisions in increasingly complex market scenarios.Moreover, the research illuminates the potential for these advanced AI models to democratize financial market analysis. By making high-level forecasting tools more accessible, a wider range of market participants, including smaller investors and firms, can benefit from insights that were previously available only to larger entities with extensive resources. This democratization could lead to a more inclusive financial environment, where diverse players have the tools to make strategic decisions based on robust, data-driven insights.In addition, the exploration into GAI in financial forecasting opens up avenues for interdisciplinary collaboration. The convergence of financial expertise, data science, and artificial intelligence could foster innovative approaches and solutions, expanding the horizons of what is possible in market analysis and prediction.

**References**
[1]    Cheng Xuejun. Artificial intelligence is deeply involved in consumer finance: motivations, risks

and prevention and control [J]. "Journal of Shenzhen University" (Humanities and Social Sciences Edition), 2021, 38(3): 67-76.
[2] Ferreira FGDC, Gandomi AH, Cardoso RT N. Artificial intelligence applied to stock market trading: a review[J]. IEEE Access, 2021, 9: 30898-30917.
[3] Milana C, Ashta A. Artificial intelligence techniques in finance and financial markets: a survey of the literature[ J]. Strategic Change, 2021, 30(3): 189-209.
[4] Lee J. Access to finance for artificial intelligence regulation in the financial services industry[ J]. European Business Organization Law Review, 2020, 21(4): 731-757.
[5] Goodell JW, Kumar S, Lim WM, et al. Artificial intelligence and machine learning in finance: Identifying foundations, themes, and research clusters from bibliometric analysis[ J]. Journal of Behavioral and Experimental Finance, 2021, 32: 100577.
[6] Liu, Bo, et al. "Integration and Performance Analysis of Artificial Intelligence and Computer Vision Based on Deep Learning Algorithms." *arXiv preprint arXiv:2312.12872* (2023).
[7] Che, Chang, et al. "Deep learning for precise robot position prediction in logistics." *Journal of Theory and Practice of Engineering Science* 3.10 (2023): 36-41.
[8] Hu, Hao, et al. "Casting product image data for quality inspection with xception and data augmentation." *Journal of Theory and Practice of Engineering Science* 3.10 (2023): 42-46.
[9] Yu, Liqiang, et al. "Semantic Similarity Matching for Patent Documents Using Ensemble BERT-related Model and Novel Text Processing Method." *arXiv preprint arXiv:2401.06782* (2024).
[10] Che, Chang, et al. "Advancing Cancer Document Classification with R andom Forest." *Academic Journal of Science and Technology* 8.1 (2023): 278-280.
[11] Lin, Qunwei, et al. "A Comprehensive Study on Early Alzheimer's Disease Detection through Advanced Machine Learning Techniques on MRI Data." *Academic Journal of Science and Technology* 8.1 (2023): 281-285.
[12] Huang, Jiaxin, et al. "Enhancing Essay Scoring with Adversarial Weights Perturbation and Metric-specific AttentionPooling." *2023 International Conference on Information Network and Computer Communications (INCC)*. IEEE, 2023.
[13] Che, Chang, et al. "Enhancing Multimodal Understanding with CLIP-Based Image-to-Text Transformation." *Proceedings of the 2023 6th International Conference on Big Data Technologies*. 2023.
[14] Huang, Zengyi, et al. "Research on Generative Artificial Intelligence for Virtual Financial Robo-Advisor." *Academic Journal of Science and Technology* 10.1 (2024): 74-80.
[15] Huang, Zengyi, et al. "Application of Machine Learning-Based K-Means Clustering for Financial Fraud Detection." *Academic Journal of Science and Technology* 10.1 (2024): 33-39.
[16] Yu, Liqiang, et al. "Stochastic Analysis of Touch-Tone Frequency Recognition in Two-Way Radio Systems for Dialed Telephone Number Identification." *arXiv preprint arXiv:2403.15418* (2024).